\documentstyle[psfig, rotate]{aa}

\begin{document}

\topmargin 0mm
\title{Cerenkov Line-like Radiation, The Extended and Improved Formulae
System}

   \author{J.H.You\inst{1}, Y.D. Xu\inst{1}, D.B.Liu\inst{1}, J.R.Shi\inst{2},
    G.X.Jin\inst{1}}

\institute{
{\inst{1} Department of Applied Physics, Shanghai Jiao Tong University,
Shanghai, 200030, China} \\
{\inst{2} Beijing Astronomical Observatory, Chinese Academy of Sciences,
Beijing 100012, China.} \\}

\date{Received ; accepted }
\thesaurus{(02.01.2; 02.12.3; 13.25.3)}

\offprints{Jun-han You (jhyou@online.sh.cn)}

\maketitle
\markboth
{You et al.: Cerenkov line-like radiation, the extended and improved
formulae system}
{You et al.: Cerenkov line-like radiation, the extended and improved
formulae system}

\begin{abstract}

You \& Cheng (1980) argued that, for relativistic electrons
moving through a dense gas, the Cerenkov effect will produce peculiar atomic
and/or molecular emission lines--Cerenkov lines. They presented a series of
formulae to describe the new line-mechanism. Elegant experimental confirmation
has been obtained by Xu. et al. in the laboratory which definitely verified
the existence of Cerenkov lines. Owing to the potential importance in high
energy astrophysics, in this paper we give a more detailed physical
discussion of the emission mechanisms and improve the previous formulae
system into a form which is more convenient for astrophysical application.
Specifically, the extended formulae in this paper can be applied to other
species of atoms and/or ions rather than hydrogen as in the previous paper
, and also to X-ray astronomy. They can also be used for the calculation of
 a nonuniform
plane-parallel slab of the emissive gas.

\end{abstract}

\begin{keywords}
Accretion, accretion disks -- Line: profiles -- X-rays: general 
\end{keywords}

\section{Introduction}

~~Early in the 1980's, You \& Cheng (1980) mentioned that, when relativistic
electrons pass through a dense gas region, the radiation produced by the
Cerenkov effect will be concentrated into a narrow waveband $\Delta\lambda$,
very near to the intrinsic atomic or molecular wavelength position $\lambda
_{lu}$ ($u,l$ denote the corresponding upper and lower levels respectively).
Therefore it looks more like an atomic emission line (for gas composed of
atoms) and/or molecular line (for molecular gas), rather than a continuum.
They called it the ``Cerenkov Emission Line''. Later, they presented a
series of formulae to describe the properties of the peculiar Cerenkov line
(You,Kiang and Chengs, 1984, 1986). Elegant experimental confirmation of Cerenkov line
in $O_{2}$, $Br_{2}$ gas, and $Na$ vapor using a $^{90}Sr$  $\beta$-ray source
has been obtained in the laboratory (Xu et al. 1981, 1988, 1989). By use of
the fast coincidence technique, they found the line emission at the expected
directions, wavelengths and plane of polarization. It has been emphasized (You
et al. 1986) that the Cerenkov emission line is not a real line in the precise
meaning due to the following features: (I) it is rather broad, for a dense
gas with $N_{H}\simeq 10^{15}{\rm cm^{-3}}$, $T\simeq 10^{4}{\rm K}$, the calculated
linewidth of Cerenkov $L_{y\alpha}$ is $\Delta\lambda\simeq 1 - 10\AA$;
(II) generally, the line profile is asymmetric, being steep on the blue side
and flat on the red side; (III) its peak is not precisely at $\lambda_{lu}$,
but slightly red-shifted, we call this the ``Cerenkov line redshift'', so as to
distinguish it from other types of redshift mechanisms (Doppler, gravitational,
Compton etc.); (IV) it is polarized if the relativistic electrons have an
anisotropic velocity distribution.

The schematic profile of the Cerenkov line $I^{c}_{\lambda}\sim \lambda$ is
shown in Fig.1, where the profile of the normal emission line
$I^{s}_{\lambda}\sim \lambda$ produced by a spontaneous transition process
 (recombination-cascade and/or collisional excitation) is also presented for
 comparison.


Here, we emphasize the special importance of point (III). It is the Cerenkov
line-redshift which strengthens the emergent intensity of the
Cerenkov emission line. The reason is easy to understand. For an optically
thick dense gas, the emergent line flux is determined both by the emission
and the absorption.The absorption mechanisms for the normal lines
(recombination-cascade and collisional excitation)
and for the Cerenkov lines are extremely different.
The line-intensity $I^{s}$ of a normal line is greatly
weakened by the large resonance absorption $k_{lu}(\lambda_{lu})$ due to the
fact that the normal line is located at the position of the intrinsic
wavelength $\lambda=\lambda_{lu}$ (see Fig.1). where $k_{lu}(\lambda_{lu})$
is very large, $k_{lu}(\lambda_{lu})\rightarrow\infty $ (In the extreme
case of a very dense gas, the emergent flux has a continuum with black body
spectrum, and the line intensity $I^{s}\simeq 0$). However, the Cerenkov line,
located at $\lambda > \lambda_{lu}$ because of the small ``Cerenkov
line-redshift'' (generally, $\Delta Z^{c}\equiv\frac {\Delta \lambda_{p}}
{\lambda _{lu}}\simeq 10^{-3}$, or the apparent ``velocity'' V a few
hundred {\rm km/sec}) can avoid the strong line absorption $k_{lu}$ because $k_{lu}
(\lambda > \lambda_{lu})\rightarrow 0$. Therefore, the Cerenkov line intensity
$I^{c}$ is only affected by a very small photoelectric absorption $k_{bf}$,
and $k_{bf}\ll k_{lu}(\lambda_{lu})$, which means that the photons of
the Cerenkov line can easily escape from deep inside a dense gas cloud. The
depth scale is $L\simeq 1/k_{bf}\gg 1/k_{lu}(\lambda_{lu})$, causing a strong
emergent intensity $I^{c}$ only if the density of fast electrons $N_{e}$
is large enough. In other words, the dense gas appears to be more
``transparent'' for the Cerenkov lines than for the normal lines,
which makes it possible
for the Cerenkov line mechanism to dominate over the normal
line emission in some physical situations. 

Such a new line emission mechanism has a potential importance for 
high energy astrophysics, particularly in X-ray astronomy. Solar flares,
QSOs and other AGNs etc. would
be the first candidates. It is known that there definitely exist abundant
relativistic electrons and dense gas regions in these objects, which provide
the right conditions for producing the Cerenkov line-like radiation. You
et al. tried to explain a series of particular observation characteristics
of the broad lines of QSOs and Seyfert1 galaxies by use of the new mechanism.
Moderate successes have been obtained,
at least qualitatively, e.g. (1) the anomalous
line-intensity ratios (the steep Balmer decrement, the small ratio
$L_{y\alpha}/H_{\beta}$, etc.)(You \& Cheng 1987); (2) the small redshift
and red-asymmetry of the broad $H_{\beta}$ line with respect to CIV
$\lambda 1549$ (Cheng et al. 1990); (3) the different lag times $\tau$ of the
high-ionized lines, e.g. CIV and the low-ionized lines, e.g. MgII
$\lambda 2798$, $H_{\beta}$, with respect to the flare of the continuum from
the central source. The progress in the study of QSOs and AGNs in recent years,
both in theories and observations mean that we can now make qualitative
statements on the contribution of the Cerenkov line mechanism to the broad
lines of AGNs.
The latest observations which seem favorable to that Cerenkov line mechanism
were provided by ASCA observations of Seyfert1 galaxies, which show an iron
$K_{\alpha}$ broad line with a markedly asymmetric profiles (Tanaka, et al. 1995, K. Nandra, et al.
1997, Turner, et al. 1997). The very steep blue side and strong red wing of
broad $K_{\alpha}$ line might be indicative of Cerenkov line which encourages
us to reexamine and improve the formulae system of the Cerenkov line
mechanism given in our previous papers (You et al. 1984, 1986), in a form
which is convenient for astrophysical applications.
In particular, we hope that the extended formulae can be applied to
other species of multi-electron complex atoms and/or ions rather than
hydrogen or hydrogenic atoms
as in the previous paper, which are
important for the study of the broad lines of AGNs, particularly in the X-ray
band. Another extension concerns the discussion on the Cerenkov line
redshift in Sect. 2, where we give a generalized formula of
the Cerenkov line redshift
and its simplified forms in the limiting cases of high and low gas-densities
respectively. We find that the conclusion of asymmetry of the line-profile is
correct only for a dense gas whose density is larger than a given critical
value $N^{cr.}$. When $N\ll N^{cr.}$, the profile becomes more symmetric. Finally,
our new formulae system is extended to the case of
a non-uniform plane-parallel slab of
emission gas ($N=N(s), T=T(s)$ etc., where $s$ is the inner distance to
the surface of
the slab), which is also important in practice.

\section{Basic Formulae}

~~The CGSE system of units will be used throughout. This means, in particular,
that all wavelengths in the following formulae will be in centimeters
rather than $\AA$. The photon energy will be in {\rm ergs}
rather than {\rm KeV}.
\subsection{The refractive index $n_{\lambda}$ and the extinction coefficient
$\kappa_{\lambda}$ }

\subsubsection{Formulae for hydrogen or hydrogen-like gas}

~~The essential point of the calculation of the spectrum of Cerenkov
radiation is the evaluation of the refractive index $n$ of
the gaseous medium. This is easy
to understand qualitatively from the necessary condition for producing
Cerenkov radiation, $V>c/n_{\lambda}$. At a given wavelength 
$\lambda$, the larger the index $n$, the easier the condition
$V>c/n_{\lambda}$ to be satisfied, and the stronger the Cerenkov radiation
at $\lambda$ will be. Therefore, in order to get the theoretical spectrum
of the Cerenkov radiation, it is necessary to calculate the refractive index
$n_{\lambda}$ and its dependence on $\lambda$, i.e. the dispersion curve
$n_{\lambda}\sim \lambda $. For a gaseous medium, the calculation of the
$n_{\lambda}\sim \lambda$ curve is easy to do. Because our main interest
is in the calculation of $n_{\lambda}$ in the vicinity of $\lambda_{lu}$, we
must use the rigorous formula:
\begin{equation}
\frac {\tilde{n}^{2}-1}{\tilde{n}^{2}+2}=\frac {4\pi}{3}N\alpha,
\end{equation}
where
\begin{equation}
\tilde{n}=n-i\kappa
\end{equation}
$\tilde{n}$ is the complex refractive index, the real part $n$ is the
refractive index of the gas, the imaginary part $\kappa$ is the extinction
coefficient. $N$ is the number density of the atoms, and $\alpha$ is the
polarizability per
atom. When the atoms are distributed over various energy levels with density
$N_{a}$ in level $a$, $N\alpha$ will be replaced by
$\sum_{a}N_{a}\alpha_{a}$. According to quantum theory, the atomic
polarizability per atom in level $a$ is given as (e.g. Handbook of Physics 
ed. Condon and Odishaw, 1958):
\begin{equation}
\alpha_{a}(\omega)=\frac{e^{2}}{m}\sum\limits_{a\not=b}\frac{f_{ab}}
{(\omega^{2}_{ab}-\omega^{2})+i\Gamma_{ab}\omega}
\end{equation}
where $e$ and $m$ are the charge and the mass of an electron, $f_{ab}$ and
$\Gamma_{ab}$ are the oscillation strength and the damping constant for the
atomic line of frequency $\omega_{ab}$. Substituting Eq.(3) into Eq.(1), we
get:
\begin{equation}
N\alpha=\sum\limits_{a}N_{a}\alpha_{a}=\sum\limits_{a}N_{a}\sum\limits_{b\not=a}
\frac{e^{2}}{2\pi m}\frac {f_{ab}}{2\pi(\nu^{2}_{ab}-\nu^{2})+i\Gamma_{ab}
\nu}
\end{equation}

Because we are concerned with the neighborhood of a given atomic
line $\lambda_{lu}$, or the intrinsic frequency $\nu_{lu}$, $\nu\simeq
\nu_{lu}$ ( the subscripts $u$ and $l$ denote the upper and lower-levels
corresponding to the intrinsic frequency $\nu_{lu}$), it follows that we
only need to keep the two largest terms in the above summation, Eq.(4) becomes:
\begin{equation}
N\alpha\simeq N_{l}\alpha_{lu}+N_{u}\alpha_{ul}
\end{equation}
where
\begin{equation}
\alpha_{lu}=\frac {e^{2}}{2\pi m}\frac {f_{lu}}{2\pi(\nu^{2}_{lu}-\nu^{2})
+i\Gamma_{lu}\nu}
\end{equation}
and $\alpha_{ul}$ is obtained from Eq.(6) by replacing $f_{lu}$ by $f_{ul}$.
Note that $f_{lu}$ and $f_{ul}$ represent the absorption and emission
oscillation strengths respectively, corresponding to the pair of energy
levels $(l,u)$ of a given atom. Therefore, $f_{lu}$ and $f_{ul}$ are related
to each other through the statistical weights $g_{l}$ and $g_{u}$, $g_{l}
f_{lu}=-g_{u}f_{ul}$. The absorption oscillation strength $f_{lu}$ is related
with Einstein's spontaneous emission coefficient $A_{ul}$ as (Bethe
\& Salpeter 1957):
\begin{equation}
f_{lu}=\frac{mc^{3}}{8\pi^{2}e^{2}\nu^{2}_{lu}}\left ( \frac {g_{u}}
{g_{l}}\right ) A_{ul}
\end{equation}
and the damping constant $\Gamma_{lu}$ is also related with Einstein's
coefficient $A_{ul}$ as:
\begin{equation}
\Gamma_{lu}=\Gamma_{l}+\Gamma_{u}=\sum\limits_{i(<l)}A_{li}+\sum\limits
_{j(<u)}A_{uj}
\end{equation}
Using these relations, Eq. (1) becomes:
\begin{equation}
\frac{\tilde{n}^{2}-1}{\tilde{n}^{2}+2}=\frac{b}{z+ig}
\end{equation}
with
\begin{displaymath}
b \equiv \frac {c^{3}}{12\pi^{2}}\nu^{-2}_{lu}A_{ul}g_{u}\left (
\frac {N_{l}}{g_{l}}-\frac {N_{u}}{g_{u}}\right )
\end{displaymath}
\begin{equation}
z \equiv 2\pi\left ( \nu^{2}_{lu}-\nu^{2}\right )
\end{equation}
\begin{displaymath}
g \equiv \Gamma_{lu}\nu
\end{displaymath}

From Eq. (9) and Eq. (2), we obtain the refractive index $n_{\nu}$ and
the extinction coefficient $\kappa_{\nu}$ as:
\begin{displaymath}
n^{2}_{\nu} =\left [\left (A^{2}+9b^{2}g^{2}\right )^{1/2}+A\right ]/2B
\end{displaymath}
\begin{equation}
\kappa^{2}_{\nu} =\left [\left (A^{2}+9b^{2}g^{2}\right )^{1/2}-A\right ]/2B
\end{equation}
where
\begin{displaymath}
A \equiv (z+2b)(z-b)+g^{2}
\end{displaymath}
\begin{displaymath}
B \equiv (z-b)^{2}+g^{2}
\end{displaymath}

Eq. (11) is the $n^{2}_{\nu}\sim\nu$ dispersion formula, which we need.
The $n^{2}_{\nu}\sim\nu$ dispersion curve can be easily obtained by the
digital calculation by use of Eq. (11).

However, as a good approximation, Eq. (11) can be replaced by a simple analytic
formula as follows. For convenience in comparing with the observations, it is
better to replace the frequency $\nu$ in Eq. (9) by wavelength $\lambda$,
let $\Delta\lambda\equiv\lambda-\lambda_{lu}$ represent the wavelength
displacement, and let
\begin{equation}
y\equiv\frac{\lambda-\lambda_{lu}}{\lambda_{lu}}=\frac {\Delta\lambda}
{\lambda_{lu}}
\end{equation}
denote the fractional displacement, $y\ll 1$ because we are always interested
in the calculation in the neighborhood of $\lambda_{lu}$, $\lambda\simeq
\lambda_{lu}$. Therefore the small dimensionless quantity $y$ will be very
useful in theoretical analysis (e.g. the expansion series by use of the small
quantity $y$). Replacing $\nu\rightarrow\lambda$ in Eq. (11), $\nu=c/\lambda
=c\lambda^{-1}_{lu}(1+y)^{-1}$, and $\nu_{lu}=c\lambda^{-1}_{lu}$.
Substituting this into Eq. (10) and Eq. (11), and keeping only the terms of
the lowest order in $y$, we get:
\begin{displaymath}
b = \frac {c}{12\pi^{2}}\lambda^{2}_{lu}A_{ul}g_{u}\left (
\frac {N_{l}}{g_{l}}-\frac {N_{u}}{g_{u}}\right )
\end{displaymath}
\begin{equation}
z = 4\pi c^{2}\lambda^{-2}_{lu}y
\end{equation}
\begin{displaymath}
g = c\lambda^{-1}_{lu}\Gamma_{lu}
\end{displaymath}

Therefore, among the three quantities $b$, $g $ and $z$, only $z$ is
$\Delta\lambda$-dependent. Because the Cerenkov line emission is not
located in the exact position $\lambda=\lambda_{lu}$, it has a small redshift,
so $y$ is not an indefinitely small quantity ($y\not\rightarrow 0$, in fact,
at $y=0$, i.e. $\lambda=\lambda_{lu}$, the Cerenkov radiation disappears).
In the actually effective range of $y$, we always have:
\begin{equation}
 g\ll z, ~~~~b\ll z
\end{equation}

For example, for $Ly\alpha$, $l=1, u=2, \lambda_{12}=1.216\times 10^{-5}
{\rm cm},\Gamma_{12}=A_{21}=6.25\times 10^{8}{\rm sec^{-1}}$.
Inserting these value in
Eq. (13), we find that $g\ll z$ whenever $\Delta\lambda\geq2.0\times 10^{-4}
\AA$ (equivalently, $y\geq 10^{-7}$). On the other hand, under ordinary
physical conditions, we can safely assume $N_{2}\ll N_{1}\sim N$. Then even
for $N_{1}$ as high as $N_{1}\simeq N\simeq 10^{17}{\rm cm^{-3}}$,
$b\ll z$ will be true
whenever $\Delta\lambda$ is greater than $1.1\times 10^{-3}\AA$ (or $y\geq
10^{-6}$). A smaller $N$ will give a still lower limit of $\Delta\lambda$.
Therefore, the inequalities (14) always hold in the actually effective
Cerenkov line-width ($\Delta\lambda\simeq 1 - 10 \AA $). For the other lines,
we obtain the same conclusion (14), under similar considerations. Thus we
need only retain the terms of the lowest order of the small quantities
$g/z$ and $b/z$ in Eq. (11), and we obtain:
\begin{displaymath}
n^{2} - 1\simeq 3\frac {b}{z}
\end{displaymath}
\begin{equation}
\kappa\simeq \frac{2}{3}\left ( \frac {b}{z}\right )\left ( \frac {g}{z}\right )
\end{equation}

Substituting Eq. (13) into Eq. (15), finally we have the simplified
approximate formulae:
\begin{displaymath}
n^{2} - 1=\frac {1}{16\pi^{3}c}\lambda^{4}_{lu}A_{ul}g_{u}\left (
\frac {N_{l}}{g_{l}}-\frac {N_{u}}{g_{u}}\right ) y^{-1}
\end{displaymath}
\begin{equation}
\kappa = \frac {1}{128\pi^{4}c^{2}}\lambda^{5}_{lu}\Gamma_{lu}A_{ul}g_{u}
\left ( \frac {N_{l}}{g_{l}}-\frac {N_{u}}{g_{u}}\right ) y^{-2}
\end{equation}

Eqs.(16) are the approximate analytic formulae for $n$ and $\kappa$
respectively. From Eq. (16) we see that $n_{\lambda}$ varies as $(n^{2}-1)
\propto y^{-1}\propto \Delta\lambda^{-1}$. We shall see below that the
Cerenkov spectral emissivity varies approximately as $J_{\lambda}\propto
n^{2}-1$, so $J_{\lambda}\propto \Delta\lambda^{-1}$. However, the
absorption varies as $\kappa_{\lambda}\propto y^{-2}\propto \Delta\lambda^{-2}$,
i.e. the absorption decreases with $\Delta\lambda$ more rapidly than the
emissivity. That is why we have a net Cerenkov radiation at the position
$\lambda=\lambda_{lu}+\Delta\lambda$, if $\Delta\lambda$ is not approaching
to zero.

The calculated $n_{\lambda}\sim \lambda$ and $\kappa_{\lambda}\sim \lambda$
curves are shown in Fig. 2. The shaded region (in Fig. 2) represents a narrow
waveband $\Delta\lambda$, very near to the position of intrinsic wavelength
$\lambda_{lu}$, but with small redshift, where the extinction $\kappa_{\lambda}
\simeq 0$, and the net Cerenkov radiation survives. The waveband $\Delta
\lambda$ is so narrow(e.g. a few $\AA$), that  looks more like a
line-emission, rather than a continuum.


\subsubsection{Extended formulae for gas of complex atoms(ions) in optical waveband}

~~We emphasize that our derivation from Eq. (4) to Eq. (16) is limited to
gas composed of hydrogen and/or hydrogen-like atoms(ions), e.g. $Na, He^{+},
Fe^{+25}$, etc. However, a necessary extention to other multi-electron atoms
(ions) can easily  be done when we notice that Eq. (3) represents the
contribution of an electron lying in the atomic energy level $a$ to the
atomic polarizability $\alpha $. Therefore, for a given emission line
of a given multi-electron atom with frequency $\nu_{lu}$,
the quantity $N\alpha $ in Eq. (1) and (5) has to be replaced by
$N(S_{l}\alpha_{lu}+S_{u}\alpha_{ul})$, i.e.
\begin{displaymath}
N\alpha\simeq N_{l}\alpha_{lu}+N_{u}\alpha_{ul}\Longrightarrow N(S_{l}\alpha
_{lu}+S_{u}\alpha_{ul})~~~~~~~~~~~(5^{\prime})
\end{displaymath}
where $N$ is the number density of complex atoms. $S_{l}$ (or $S_{u}$)
represents the actual occupation number of the electrons at the lower level
$l$ (or upper level $u$) of complex atom. Obviously we have
\begin{displaymath}
S_{l}\leq g_{l}~~~~{\rm and}~~~~~S_{u}\leq g_{u}
\end{displaymath}
where $g_{l}$ and $g_{u}$ are the degeneracy of level $l$ and $u$
respectively. From Eq. (1), ($5^{\prime}$) and equality
$g_{l}f_{lu}=-g_{u}f_{ul}$, it is easy to see that, if the levels ($l$, $u$)
have been fully occupied by electrons (`closed shell'), i.e. $S_{l}=g_{l}$
and $S_{u}=g_{u}$, then we have $N\alpha =0 $,
thus $\tilde{n}_{\nu}=n_{\nu}-i\kappa_{\nu}=1$.
In other words, for the multi-electron complex atoms, the Cerenkov
line-like emission related to frequency $\nu_{lu}$ can occur only when
the upper and/or lower levels ($l, u$) are not fully occupied, i.e.
$S_{l}<g_{l}$, and/or $S_{u}<g_{u}$ \footnote[1]{This is possible for
high temperature cosmological plasmas, where the high ionized ions exists,
e.g. for a plasma with $T\simeq 2-3\times 10^{6}K$, the most abundant
ion-species of iron is $Fe^{+16}$ for which $S_{1}=2=g_{1}$, $S_{2}=8=g_{2}$, but $S_{3}
=0<g_{3}$. If $T\simeq 10^{7}K$, both the most abundant and averaged species
 of ironions is $Fe^{+19}$, for which $S_{1}=2=g_{1}$, and
 $\overline{S_{2}}\simeq 5<g_{2}.$}.
In this case, the formulae for
quantities $Z$ and $g$ in Eq. (9) are same as in Eq. (10), but the
representation of quantity $b$ has to be changed. When
$N_{l}\alpha_{lu}+N_{u}\alpha_{ul}$
is replaced by $N(S_{l}\alpha_{lu}+S_{u}\alpha_{ul})$,
we obtain
\begin{displaymath}
b\equiv \frac {c^{3}}{12\pi^{2}}\nu^{-2}_{lu}A_{ul}g_{u}N\left(
\frac {S_{l}}{g_{l}}-\frac {S_{u}}{g_{u}} \right ) ~~~~~~~~~~~~~~~~~~
~~~~~~~(10^{\prime})
\end{displaymath}
or
\begin{displaymath}
b\equiv\frac{c}{12\pi^{2}}\lambda^{2}_{lu}A_{ul}g_{u}N\left (
\frac{S_{l}}{g_{l}}-\frac{S_{u}}{g_{u}}\right )~~~~~~~~~~~~~~~~~~~
~~~~~~(13^{\prime})
\end{displaymath}
Thus the final forms of the Eq. (16) for gas composed of
multi-electron complex atoms (ions) become
\begin{displaymath}
n^{2}-1=\frac {1}{16\pi^{3}c}\lambda^{4}_{lu}A_{ul}g_{u}N\left (
\frac {S_{l}}{g_{l}}-\frac {S_{u}}{g_{u}}\right ) y^{-1}
\end{displaymath}
\begin{displaymath}
\kappa_{\lambda}=\frac{1}{128\pi^{4}c^{2}}\lambda^{5}_{lu}\Gamma_{lu}
A_{ul}g_{u}N\left ( \frac {S_{l}}{g_{l}}-\frac {S_{u}}{g_{u}}\right )
y^{-2}~~~~~~~~(16^{\prime})
\end{displaymath}

In the following sections of this paper, all formulae from Eq. (17) to
Eq. (42) are used for hydrogen and hydrogen-like gas both for simplicity and
practicality because hydrogen is the most abundant element in the cosmological
plasmas. However, most of these equations are valid for the multi-electron
complex atoms only if the following replacement is made:
\begin{displaymath}
\left (\frac {N_{l}}{g_{l}}-\frac {N_{u}}{g_{u}}\right )\Longrightarrow
N\left (\frac {S_{l}}{g_{l}}-\frac {S_{u}}{g_{u}}\right )~~~~~~~~~~~~~~~~~~~~~
~~(5^{\prime})
\end{displaymath}

\subsubsection{Extended formulae for complex atoms(ions) gas in the X-ray
waveband}

~~Another important extension, which we would like to emphasize in this paper,
is related to the possible application of Cerenkov line-like emission in 
X-ray astronomy because the waveband of some Cerenkov lines of complex atoms
(ions), particularly the heavy elements, is in the X-ray range. A possible
application of the Cerenkov line mechanism in the X-ray band may be a new
explanation for the origin of the broad $K_{\alpha}\sim 6.4 {\rm KeV}$
line of Fe ions observed in Seyfert AGNs in recent years. For convenience in
comparing with the observations in the X-ray waveband, it is better
to replace the wavelength $\lambda$ in the formulae given above by the
frequency $\nu$, or equivalently, by the photon energy $\epsilon=h\nu$. This
is easy to do without any remarkable changes in above formulae. We notice
that the dimensionless small quantity $y$ in Eq. (12) can also be written as
\begin{displaymath}
y\equiv \frac {\lambda-\lambda_{lu}}{\lambda_{lu}}\equiv\frac{\Delta\lambda}
{\lambda_{lu}}=-\frac {\Delta\nu}{\nu_{lu}}=-\frac {\Delta\epsilon}
{\epsilon_{lu}}~~~~~~~~~~~~~~~~~~~~~~(12^{\prime})
\end{displaymath}
where $\epsilon_{lu}\equiv h\nu_{lu}=\epsilon_{u}-\epsilon_{l}$ represents
the energy difference of the upper and lower levels ($u, l$), and
$-\Delta\epsilon=h\nu_{lu}-h\nu=\epsilon_{lu}-\epsilon $.  i.e. the small
quantity $y$ also represents the fractional displacement of the frequency
or photon-energy. Inserting Eq. ($12^{\prime}$) into the equations given above
we obtain the following formulae which are suitable to the application to
X-ray astronomy:
\begin{displaymath}
n^{2}_{\nu}-1=\frac{c^{3}h^{4}}{16\pi^{3}}\epsilon^{-4}_{lu}A_{ul}
g_{u}N\left ( \frac{S_{l}}{g_{l}}-\frac {S_{u}}{g_{u}}\right ) y^{-1}
\end{displaymath}
\begin{displaymath}
\kappa_{\nu}=\frac {c^{3}h^{4}}{128\pi^{4}}\epsilon^{-5}_{lu}\Gamma_{lu}
A_{ul}g_{u} N \left ( \frac {S_{l}}{g_{l}}-\frac {S_{u}}{g_{u}} \right )
y^{-2}~~~~~~~~~~~~(16^{\prime\prime})
\end{displaymath}

\subsection{The Cerenkov spectral emissivity $J^{c}_{\lambda}(J^{c}_{y})$
and the line width $\Delta\lambda^{c}_{\rm lim}(y_{\rm lim})$ }
\subsubsection{Formulae for hydrogen or hydrogen-like gas}

~~The Cerenkov spectral emissivity can be derived from the dispersion curve
$n_{\lambda}\sim \lambda$ given above. It is known from the theory of 
Cerenkov radiation that the power emitted in a frequency interval ($\nu,
\nu+d\nu$) by an electron moving with velocity ($\beta=V/c$) is $P_{\nu}
d\nu=(4\pi^{2}e^{2}\beta\nu/c)\left ( 1-\frac {1}{n^{2}_{\nu}\beta^{2}}
\right ) d\nu$. Let $N(\gamma)d\gamma$ be the number density of fast
electrons in the energy interval ($\gamma, \gamma+d\gamma$), ($\gamma=\frac
{1}{\sqrt {1-\beta^{2}}}=mc^{2}/m_{0}c^{2}$ is the Lorentz factor, which
represents the dimensionless energy of the electron), then the power emitted
in interval ($\nu, \nu+d\nu$) by these electrons is $N(\gamma)d\gamma
P_{\nu}d\nu$. For an isotropic velocity distribution of the relativistic
electrons as in normal astrophysical conditions, the Cerenkov radiation
will also be isotropic(the definite angular distribution of the Cerenkov emission
disappears when the relativistic electrons have an isotropic distribution
of velocities). Hence, the spectral emissivity per unit volume and unit solid
angle is:
\begin{displaymath}
J^{c}_{\nu}d\nu=\frac {1}{4\pi}\int^{\gamma_{2}}_{\gamma_{1}}N(\gamma)
d\gamma P_{\nu}d\nu
\end{displaymath}
\begin{equation}
~~~~~~=\frac {\pi e^{2}}{c}\nu d\nu\int_{\gamma_{1}}
^{\gamma_{2}}N(\gamma)d\gamma\beta\left ( 1-\frac {1}{n^{2}\beta^{2}}\right )
\end{equation}
where, $\gamma_{1}, \gamma_{2}$ are the lower and upper limit of the energy
spectrum of the relativistic electrons respectively. For the relativistic
electrons, we always have $\beta\simeq1$, $\gamma\gg 1$, so $\beta^{-2}
\simeq 1+\gamma^{-2}$, $\beta\simeq 1-\frac{1}{2}\gamma^{-2}$. Also we notice
that in the actual effective emission range, the actual refractive index $n$
of a gas is not far from unity, $n\sim 1$ (see Eq. (16)). Therefore:
\begin{displaymath}
\int_{\gamma_{1}}^{\gamma_{2}}N(\gamma)\left (
1-\frac {1}{n^{2}\beta^{2}}\right )\beta d\gamma
\simeq\int_{\gamma_{1}}^{\gamma_{2}}\left (n^{2}-1-\gamma^{-2}\right )
N(\gamma)d\gamma
\end{displaymath}
\begin{displaymath}
~~~~~~~~~~~~~~~\simeq\left ( n^{2}-1-\gamma^{-2}_{c}\right ) N_{e}
\end{displaymath}
where, $N_{e}\equiv \int_{\gamma_{1}}^{\gamma_{2}}N(\gamma)d\gamma $ is the
density of the relativistic electrons, and $\gamma_{c}$ is the characteristic
energy of the electrons (or the typical energy) in a given source. The
definition of $\gamma_{c}$ is $\int_{\gamma_{1}}^{\gamma_{2}}\gamma^{-2}
N(\gamma)d\gamma\equiv \gamma^{-2}_{c}N_{e}$, so $\gamma_{1}<\gamma_{c}
<\gamma_{2}$. Hence:
\begin{equation}
J^{c}_{\nu}d\nu\simeq \frac {\pi e^{2}}{c}N_{e}\nu\left (
n^{2}-1-\gamma^{-2}_{c}\right ) d\nu
\end{equation}

Replacing $\nu$ to $\lambda$ or $y$, we have $J_{\nu}^{c}d\nu=J_{\epsilon}^{c}
d\epsilon=J_{\lambda}^{c}d\lambda=J^{c}_{y}dy$ and $dy=\lambda_{lu}^{-1}
d\lambda=\lambda^{-1}_{lu}(-c\nu^{-2}d\nu)$. Therefore:
\begin{equation}
J_{y}^{c}dy=\pi c e^{2}N_{e}\lambda^{-2}_{lu}\left ( n^{2}-1-\gamma^{-2}_{c}
\right ) dy
\end{equation}

Setting $n^{2}-1-\gamma^{-2}_{c}=0$ and inserting the expression for $n$
(Eq. (16)) into Eq. (19), we get the Cerenkov line-width $\Delta\lambda^{c}
_{\rm lim}$:
\begin{displaymath}
y_{\rm lim}\equiv\frac {\Delta\lambda^{c}_{\rm lim}}{\lambda_{lu}}=C_{0}\gamma^{2}
_{c}
\end{displaymath}
\begin{equation}
~~~~~=\frac {1}{16\pi^{3}c}\lambda^{4}_{lu}A_{ul}g_{u}\left( \frac {N_{l}}
{g_{l}}-\frac {N_{u}}{g_{u}}\right ) \gamma^{2}_{c}
\end{equation}
where
\begin{displaymath}
C_{0}=\frac {1}{16\pi^{3}c}\lambda^{4}_{lu}A_{ul}g_{u}\left ( \frac {N_{l}}
{g_{l}}-\frac {N_{u}}{g_{u}}\right )
\end{displaymath}

The Cerenkov radiation will be cut off at the wavelength displcaement
$\Delta\lambda^{c}_{\rm lim}$. Inserting Eqs. (16) and (20) into Eq. (19), the
spectral emissivity becomes:
\begin{equation}
J^{c}_{y}dy=C_{1}N_{e}\left ( y^{-1}-y^{-1}_{\rm lim}\right ) dy
\end{equation}
where
\begin{displaymath}
C_{1}\equiv\frac{e^{2}}{16\pi^{2}}\lambda^{2}_{lu}A_{ul}g_{u}\left (
\frac {N_{l}}{g_{l}}-\frac {N_{u}}{g_{u}}\right )
\end{displaymath}

It is obvious that the emissivity $J^{c}_{\lambda}=0$, when $y=y_{\rm lim}$
(or $\Delta\lambda=\Delta\lambda^{c}_{\rm lim}$). But for small wavelength
displacement $\Delta\lambda$, $y\ll y_{\rm lim}$, we have
$J_{y}^{c}dy\propto y^{-1}\propto\Delta\lambda^{-1}$,
i.e. $J^{c}_{y}$ decreases with $\Delta\lambda$ slowly as
$J^{c}_{y}\propto\Delta\lambda^{-1}$. Therefore, in the
actual effective emission range $y\ll y_{\rm lim}$, we have a good approximation
of Eq. (21)
\begin{equation}
J^{c}_{y}dy=C_{1}N_{e}y^{-1}dy
\end{equation}

We point out that there is a remarkable difference between the Cerenkov line
emission and the usual spontaneous radiation transition, the later is
determined by the population density in the upper level $u$, $J^{s}\propto
N_{u}$. But the Cerenkov emissivity is determined by the difference in
populations between the lower and upper levels $\left( \frac {N_{l}}{g_{l}}
-\frac {N_{u}}{g_{u}}\right ) $. 

The original profile of the Cerenkov line is given by the calculated curve
$J^{c}_{y}\sim y$ $\left( or J^{c}_{\epsilon}\sim \Delta\epsilon , or J^{c}
_{\lambda}\sim \Delta\lambda \right)$, as shown in Fig. 3(a) which is
obtained from Eq. (22). However, for the small wavelength displacement
$\Delta\lambda\simeq0$, i.e. $\lambda $ is very close to $\lambda_{lu}$,
the approximate Eq. (22)
has to be replaced by the strict Eq. (11) and Eq. (17).

\subsubsection{Extended formulae of $J^{c}_{y}$ and $y^{c}_{\rm lim}$ for complex
atoms(ions) or molecules gas}

~~The basic formulae for Cerenkov spectral emissivity $J^{c}_{y}$ and maximum
line-width $y_{\rm lim}$, Eq. (20), Eq. (21) and Eq. (22) can be easily
extended to the multi-electron complex atoms(ions) or molecular gas by use of
the same simple replacement given by Eq. ($5^{\prime}$) in Sect. 2.1.2, i.e.
$N_{l}\rightarrow NS_{l}$, $N_{u}\rightarrow NS_{u}$. This is sufficient for
the Cerenkov line of complex atoms in the optical band. However, if the Cerenkov
line is located in the X-ray band, e.g. the Cerenkov Fe $K_{\alpha}$ line at
$\sim 6.4 {\rm KeV}$, it is more convenient to replace the wavelength $\lambda$ in
Eqs. (20), (21) and (22) by the photon energy $\epsilon=h\nu$, $\lambda_{lu}
\rightarrow \epsilon_{lu}=\frac {hc}{\lambda_{lu}}$. This is easy to do
because of the simple relation $J_{\lambda}^{c}d\lambda=J_{y}^{c}dy
=J_{\epsilon}^{c}d\epsilon $ and $y=\frac {\lambda-\lambda_{lu}}{\lambda_{lu}}
=\frac {\nu_{lu}-\nu}{\nu_{lu}}=\frac {\epsilon_{lu}-\epsilon}{\epsilon_{lu}}$.
Therefore, the Cerenkov spectral emissivity and line-width of complex atoms
in X-ray band are still given by Eq. (20), (21) and (22) respectively, but
the parameters are changed as
\begin{displaymath}
C_{0}\equiv \frac {c^{3}h^{4}}{16\pi^{3}}\epsilon^{-4}_{lu}A_{ul}g_{u}
N\left ( \frac {S_{l}}{g_{l}}-\frac {S_{u}}{g_{u}}\right )~~~~~~~~~~~~~~
~~~~(20^{\prime})
\end{displaymath}
and
\begin{displaymath}
C_{1}\equiv \frac {e^{2}c^{2}h^{2}}{16\pi^{2}}\epsilon^{-2}_{lu}A_{ul}
g_{u}N\left ( \frac {S_{l}}{g_{l}}-\frac {S_{u}}{g_{u}}\right )~~~~~~~~~~~~~~
~~~~(21^{\prime})
\end{displaymath}

We emphasize again that $\epsilon_{lu}=\epsilon_{u}-\epsilon_{l}$ in Eq. ($20
^{\prime}$) and ($21^{\prime}$) is in unit of {\rm ergs} in CGSE system,
rather than in {\rm KeV} as usually used in the X-ray astronomy.

\subsection{The absorption coefficient}
\subsubsection{Formulae for hydrogen and hydrogen-like gas}

~~For an optically thick dense gas for which the Cerenkov line mechanism is
efficient, the final emergent intensity $I^{c}_{\lambda}$ is determined by
the competition between the emission $J^{c}_{\lambda}$ and the absorption
$k_{\lambda}$. Therefore it is necessary to consider the absorption of the
gas at $\lambda\simeq\lambda_{lu}$. For the optical and X-ray wavebands,
there are two main absorption mechanisms that are relevant to the Cerenkov
line emission. One is the line absorption $k_{lu}$ in the vicinities of
atomic lines, directly related to the extinction coefficient $\kappa$ given
in Eqs. (11) and (16) by the relation $k_{lu}=4\pi \kappa/\lambda$.
Another is the photoelectric absorption $k_{bf}$, the free-free absorption
$k_{ff}$ is very small in the optical or X-ray band in which we are mainly
interested, and can be neglected. Thus, for us, in the dust-free case,
the total absorption is:
\begin{equation}
k=k_{lu}+k_{bf}
\end{equation}
$k_{lu}$ can be obtained from the well known formula in molecular optics,
$k_{lu}=\frac {4\pi}{\lambda}\kappa_{\lambda}=\frac {4\pi}{\lambda_{lu}}(1-y)
\kappa_{\lambda}\simeq\frac {4\pi}{\lambda_{lu}}\kappa_{\lambda}$. Using
Eq. (16) and retaining the lowest order of $y$, we have:
\begin{equation}
k_{lu}=\frac{1}{32\pi^{3}c^{2}}\lambda^{4}_{lu}A_{ul}\Gamma_{lu}g_{u}
\left (\frac{N_{l}}{g_{l}}-\frac{N_{u}}{g_{u}}\right )y^{-2}=C_{2}y^{-2}
\end{equation}
where
\begin{displaymath}
C_{2}\equiv\frac {1}{32\pi^{3}c^{2}}\lambda^{4}_{lu}A_{ul}\Gamma_{lu}g_{u}
\left (\frac{N_{l}}{g_{l}}-\frac{N_{u}}{g_{u}}\right )
\end{displaymath}
Therefore the resonance absorption $k_{lu}\propto\Delta\lambda^{-2}$
decreases rapidly with $\Delta\lambda$.

Here, we point out that $k_{lu}$ can be obtained in another way which is more
familiar to astronomers. The well known formula for line absorption is:
\begin{displaymath}
k_{lu}=\frac{c^{2}N_{l}}{8\pi\nu^{2}}\left ( \frac {g_{u}}{g_{l}}\right )
\left ( 1-\frac {g_{l}N_{u}}{g_{u}N_{l}}\right )A_{ul}\varphi_{ul}(\nu)
\end{displaymath}
where, $\varphi_{ul}(\nu)$ is the Lorentz profile factor
\begin{displaymath}
\varphi_{ul}(\nu)=\frac {\Gamma_{lu}/4\pi^{2}}{(\nu-\nu_{lu})^{2}+(\Gamma
_{lu}/4\pi)^{2}}\simeq \frac {\Gamma_{lu}}{4\pi^{2}(\nu-\nu_{lu})^{2}}
\end{displaymath}
Combining these two expressions, we obtain Eq. (24) again.

The photoionization absorption coefficient is
$k_{bf}=\sum_{s}N_{s}\sigma_{bf}(s)$,
where the summation extends over all levels for which the photoionization
potential is less than incident photon energy $h\nu$, $I_{s}\leq h\nu$.
In normal astrophysical conditions, the most important photoionization
absorber is the neutral hydrogen due to its great abundance. Therefore, in
optical wavebands we only consider the absorption by neutral hydrogen in
the calculation of $k_{bf}$ ($H_{e}$ is the next candidate in the detailed
calculation). The photoelectric cross section of level s of the hydrogen atoms
is:
\begin{displaymath}
\sigma_{bf}(s)=2.8\times 10^{29}\nu^{-3}s^{-5}
\end{displaymath}
or
\begin{displaymath}
\sigma_{bf}(s)=1.04\times 10^{-2}s^{-5}\lambda^{3}_{lu}(1+y)^{3}\simeq
1.04\times 10^{-2}s^{-5}\lambda^{3}_{lu}
\end{displaymath}
(we keep the lowest order of $y$) Therefore
\begin{displaymath}
k_{bf}\simeq 1.04\times 10^{-2}\lambda^{3}_{lu}\sum\limits_{s\geq p}
N_{H^{0}_{s}}s^{-5}
\end{displaymath}
\begin{equation}
~~~~~\simeq 1.04\times 10^{-2}\lambda^{3}_{lu}N_{H^{0}_{p}}p^{-5}
\end{equation}

The last approximation step in Eq. (25) means that only the absorption of
the lowest photoelectric level $s=p$ (i.e. the largest term in the summation)
is taken into consideration. The calculated $k_{\lambda}=k_{lu}+k_{bf}\sim\Delta
\lambda$ curve is shown in Fig. 3(b). Comparing Eq. (25) with Eq. (24), we see
that the line absorption $k_{lu}\propto y^{-2}$ decreases rapidly with
increasing $y$, so $k_{lu}$ is effective only in a very narrow range near to
$\lambda_{lu}$, while the photoelectric absorption $k_{bf}$ is nearly
independent of wavelength displacement $\Delta\lambda$ (or $y$). Over the
whole width of the Cerenkov line, $k_{bf}$ is actually the main absorption
that determines the integrated emergent intensity. The main effect of
$k_{lu}$ is to shift the line towords the red side of $\lambda_{lu}$ (i.e. the
Cerenkov line redshift).


\subsubsection{Extended formulae of absorption $k$ for complex atoms(ions) gas}

~~As for the complex atoms(ions), the absorption coefficient $k$ is still
given
by Eq. (23), where the line absorption $k_{lu}$ is still given by Eq. (24),
$k_{lu}=C_{2}y^{-2}$. But
the parameter $C_{2}$ is expressed as
\begin{displaymath}
C_{2}\equiv \frac {1}{32\pi^{3}c^{2}}\lambda^{4}_{lu}A_{ul}\Gamma_{ul}g_{u}
N\left ( \frac {S_{l}}{g_{l}}-\frac {S_{u}}{g_{u}}\right )~~~~~~~~~~~~~~~~
(24^{\prime})
\end{displaymath}
i.e. we use the replacement $N_{l}\rightarrow NS_{l}$, $N_{u}\rightarrow
NS_{u}$, where $N$ is the density of the multi-electron complex atoms concerned.
Futhermore, if the Cerenkov line is located in the X-ray band, $C_{2}$ is
expressed as
\begin{displaymath}
C_{2}\equiv\frac{c^{2}h^{4}}{32\pi^{3}}\epsilon^{-4}_{lu}A_{ul}\Gamma_{ul}
g_{u}N\left( \frac{S_{l}}{g_{l}}-\frac{S_{u}}{g_{u}}\right )~~~~~~~~~~~
~~~~~~~~~(24^{\prime\prime})
\end{displaymath}
where $\epsilon_{lu}$ is in {\rm ergs}. i.e. we have used the replacement
$\lambda_{lu}=\frac{hc}{\epsilon_{lu}}$ in Eq. ($24^{\prime}$) to get ($24^
{\prime\prime}$).

However, when the Cerenkov line is in the X-ray band, we must take care for the
calculation of the second kind of absorption, the photoionization $k_{bf}$,
because of the fact that the hydrogen is no longer an important species
responsible for the photoelectric absorption due to its very small
photoelectric absorption cross section in the X-ray band ($\sigma_{bf}\propto
\nu^{-3}$, see Eq. (25)), in spite of the great abundance of hydrogen in
cosmological plasma.

In fact, for a given emission line in the X-ray band,
the dominant contributors to
the photoelectric absorption are the heavy elements with high $Z$, including
the relevant multi-electron atom itself, which produces the emission
line concerned (see the frequency behaviour of the
photoelectric absorption cross section shown in Eq. (26) below, which shows
the cross section $\sigma_{bf}$ reaches a maximum at each absorption edge of
the given atom, then decreases drastically as $\propto\nu^{-3}$). For
example, for Fe $K_{\alpha}$ line at $\sim 6.4 {\rm KeV}$, the dominant
photoionization absorbers are approximately the iron atoms or ions themselves, 
rather than other elements, Ca, O, S, etc., due to the highest abundance of
$Fe$ among the heavy elements with high $Z$. Furthermore, the main energy levels
of the electron of the complex atom which are responsible for the photoelectric
absorption are the K, L, M shells of the heavy elements. For example, for Fe
$K_{\alpha}$ line $\sim 6.4 {\rm KeV}$, the most important contributors to the
photoelctric absorption are the electrons in L shell of Fe ions. The
ionization potential of the Fe K-shell is $I_{K}\sim 7.2 {\rm KeV}$, higher than
$6.4 {\rm KeV}$. Thus the electrons in the K-shell of the iron atom or ions can not be
photoionized by the incident Fe $K_{\alpha}$ photons $\sim 6.4 {\rm KeV}$.
Therefore, for Fe $K_{\alpha}$ line, we have
\begin{displaymath}
k_{bf}=\sum\limits_{s\geq 2}N(Fe)S_{s}\sigma_{bf}(s)\simeq N(Fe)S_{2}\sigma
_{bf}(2)~~~~~~~~~~~~(25^{\prime})
\end{displaymath}
where $N(Fe)$ is the density of Fe in the gas, $S_{2}$ is the occupation
number of electrons at $s=2$ energy level, $S_{2}\leq g_{2}$, $\sigma_{bf}(2)$
is the photoelectric cross section of an electron at $s=2$ level. The
last approximation in ($25^{\prime}$) means that we only keep the largest
term in the summation due to the fact that $\sigma_{bf}(2)$ is the largest
one around $\sim 6.4 {\rm KeV}$, comparing with $\sigma_{bf}(3)$, $\sigma_{bf}(4)$
$\cdots$, i.e. $\sigma_{bf}(2)\gg \sigma_{bf}(3)\gg \sigma_{bf}(4)$.
For the
Fe atom and/or ions, the hydrogen-like formula for the cross section is a good
approximation, particularly for the low-lying levels $s=2, 3$
\begin{equation}
\sigma_{bf}(\nu,s)=\frac {32\pi^{2}e^{6}R_{\infty}Z^{4}}{3\sqrt{3}h^{3}
\nu^{3}s^{5}}g_{fb}(\nu,T)
\end{equation}
Therefore $\sigma_{bf}(2)$ in Eq. ($25^{\prime}$) is (taking $Z^{eff.}=24$ for $s=2$
level, i.e. for electron in the L-shell, the effective $Z^{eff.}=24$)
\begin{displaymath}
\sigma_{bf}(2)\simeq 2.9\times 10^{33}/\nu^{3}\simeq 2.9\times 10^{33}
/\nu_{lu}^{3}~~~({\rm cm^{2}})
\end{displaymath}
Inserting to Eq. ($25^{\prime}$), we get
\begin{displaymath}
k_{bf}=8.4\times 10^{-46}N(Fe)S_{2}\epsilon^{-3}_{lu}~~~~~~~~~~~~~~~~~~~
~~~~~(25^{\prime\prime})
\end{displaymath}
where $\epsilon_{lu}\equiv\epsilon_{21}=6.4{\rm KeV}
=1.025\times 10^{-8}({\rm ergs})$ in unit {\rm ergs}.

\subsection{The emergent spectral intensity $I^{c}_{\lambda}$ (or $I^{c}_{y}$).
The line profile }

~~In this section, we present formulae describing the Cerenkov intensity and
the line profile, emergent from a plane-parallel emission slab. Similarly,
we first discuss the hydrogen or hydrogen-like gas. Using the $J^{c}_{y}$
and $k_{y}$ given above, the emergent Cerenkov line intensity $I^{c}
_{\lambda}$ from the surface of the slab can be calculated by the equation of
radiative transfer:
\begin{equation}
\frac{d}{d\tau_{\lambda}}\left [ \frac {I^{c}_{\lambda}(x)}{n^{2}_{\lambda}}
\right ] =S_{\lambda}-\frac {I^{c}_{\lambda}(x)}{n^{2}_{\lambda}}
\end{equation}
where, $d\tau_{\lambda}=k_{\lambda}ds$ is the elementary optical depth,
$S_{\lambda}\equiv J^{c}_{\lambda}/k_{\lambda}n^{2}_{\lambda}$ is the source
function.

\subsubsection{$I^{c}_{\lambda}$ (or $I^{c}_{y}$) and the line profile
of the Cerenkov line of hydrogen or hydrogen-like gas} 
Case A. The uniform plane-parallel slab.

~~For a uniform plane-parallel slab of an emitting gas with thickness $L$, the
solution of Eq. (27) gives the emergent spectral intensity $I^{c}_{\lambda}$:
\begin{equation}
I^{c}_{\lambda}=n^{2}_{\lambda}S_{\lambda}(1-e^{-\tau_{\lambda}})=\frac
{J^{c}_{\lambda}}{k_{\lambda}}\left ( 1-e^{-k_{\lambda}L}\right )
\end{equation}
where $k_{\lambda}L=\tau_{\lambda}$ is the optical thickness of the uniform
slab of the emission gas. Replacing $\lambda$ by $y\equiv\Delta\lambda/
\lambda_{lu}$, Eq. (28) can be expressed in an equivalent form:
\begin{equation}
I^{c}_{y}=\frac {J^{c}_{y}}{k_{y}}\left ( 1-e^{-k_{y}L}\right )
\end{equation}

However, we shall be particularly interested in the optically thick case,
because our main interest is in the high energy astrophysical objects, such
as QSOs, supernovae, solar flares, etc., and we know that these objects all
have compact structures in which the gas density is high, and near the
optically thick case. On the other hand, in Sec. 1, we have argued that the
Cerenkov line radiation becomes important only when the gas is dense and
optically thick for continuum, i.e. $k_{\lambda}L=(k_{lu}+k_{bf})L\gg 1$.
In the optically thick case, Eq. (29) becomes $I^{c}_{y}=J^{c}_{y}/k_{y}$ or
\begin{equation}
I^{c}_{\lambda}\simeq J^{c}_{\lambda}/k_{\lambda}
\end{equation}
where, $J^{c}_{\lambda}$, $k_{\lambda}=k_{lu}+k_{bf}$ have been given in Eqs.
(21), (24) and (25), in which the variable is $y=\Delta\lambda/\lambda_{lu}$,
rather than $\Delta\lambda$. Therefore Eq. (30) becomes:
\begin{equation}
I^{c}_{y}=\frac {J^{c}_{y}}{k_{lu}+k_{bf}}=\frac {N_{e}C_{1}\left ( y^{-1}
-y^{-1}_{\rm lim}\right )}{C_{2}y^{-2}+k_{bf}}
\end{equation}
where, $C_{1}, C_{2}$ and $k_{bf}$ are given in Eqs. (21), (24) and (25)
respectively, and $y^{c}_{\rm lim}=C_{0}\gamma^{2}_{c}$, $C_{0}$ is given
in Eq. (20).

For convenience, we re-list the constants as follows:
\begin{displaymath}
C_{0}  = \frac {1}{16\pi^{3}c}\lambda_{lu}^{4}A_{ul}g_{u}\left (
\frac {N_{l}}{g_{l}}-\frac {N_{u}}{g_{u}}\right )
\end{displaymath}
\begin{displaymath}
~~~\simeq 6.72\times 10^{-14}\lambda_{lu}^{4}A_{ul}\frac {g_{u}}{g_{l}}
NR_{l}
\end{displaymath}
\begin{displaymath}
C_{1}  = \frac {e^{2}}{16\pi^{2}}\lambda_{lu}^{2}A_{ul}g_{u}\left (
\frac {N_{l}}{g_{l}}-\frac {N_{u}}{g_{u}}\right )
\end{displaymath}
\begin{displaymath}
~~~ \simeq 1.46\times 10^{-21}\lambda_{lu}^{2}A_{ul}\frac {g_{u}}{g_{l}}
NR_{l}
\end{displaymath}
\begin{displaymath}
C_{2} = \frac {1}{32\pi^{3}c^{2}}\lambda_{lu}^{4}A_{lu}\Gamma_{lu}
g_{u}\left (
\frac {N_{l}}{g_{l}}-\frac {N_{u}}{g_{u}}\right )
\end{displaymath}
\begin{displaymath}
~~~\simeq 1.12\times 10^{-24}\lambda_{lu}^{4}A_{ul}\Gamma_{lu}
\frac {g_{u}}{g_{l}}NR_{l}
\end{displaymath}
\begin{displaymath}
k_{bf} = 1.04\times 10^{-2}\lambda^{3}_{lu}N_{H_{p}^{0}}p^{-5}
\end{displaymath}
\begin{equation}
~~~ = 1.04\times 10^{-2}\lambda_{lu}^{3}N_{H^{0}}p^{-5}R_{p}
\end{equation}
(Note that the wavelength $\lambda_{lu}$ in Eqs. (32) is in
centimeter rather than $\AA$)

The last step approximation in Eqs.(32) is due to the fact that,
in the normal conditions of a gaseous medium, at least for the lowest
levels, we have $N\simeq N_{1}\gg N_{2}\gg N_{3}\cdots$, or $N_{l}\gg N_{u}$.
So $\left ( \frac {N_{l}}{g_{l}}-\frac {N_{u}}{g_{u}}\right )\simeq\frac
{N_{l}}{g_{l}}=\frac {1}{g_{l}}R_{l}N$, $N$ is the total density of the
hydrogen or hydrogenic atom (or ion) species concerned,
e.g. $N=N(Mg^{+}), N_{H^{0}}, N(Na^{0})\cdots$ for the calculation of
Cerenkov lines of $MgII,HI,NaI,\cdots$.
$R_{l}\equiv\frac {N_{l}}{N}$ is the fractional
population in the level $l$ of the concerned atom (ion) species. Similarly,
$\sum\limits_{s\geq p}N_{H^{0}_{s}}s^{-5}\simeq R_{p}N_{H^{0}}p^{-5}$,
where $N_{H^{0}}$ is the number density of the neutral hydrogen
atom in the uniform slab, $R_{p}\equiv N_{H^{0}_{p}}/N_{H^{0}}$ is the
fractional population of the neutral hydrogen in the lowest photoelectric
level $p$.

Finally, for comparison with the observations, it is necessary to
transform $I^{c}_{y}$ in Eq. (31) into $I^{c}_{\lambda}$. Using $I^{c}_{y}
dy=I^{c}_{\lambda}d\lambda$ and $dy=\frac {1}{\lambda_{lu}}d\lambda$, we
have:
\begin{equation}
I^{c}_{\lambda}=\frac {1}{\lambda_{lu}}I^{c}_{y}
\end{equation}
If the conventional unit (${\rm ergs/cm^{2}\cdot sec\cdot str\cdot\AA}$)
is adopted
for $I^{c}_{\lambda}$, then Eq. (33) becomes $I^{c}_{\lambda}=\frac {10^{-8}}
{\lambda_{lu}}I^{c}_{y}$. By use of Eqs. (31), (32) and (33), the calculated
profile of the Cerenkov line $I^{c}_{\lambda}\sim \Delta\lambda$ in the
optically thick case is shown in Fig. 3(c), which is broad (line-width
$\Delta\lambda_{\rm lim}=\lambda_{lu}y_{\rm lim}=C_{0}\lambda_{lu}\gamma^{2}_{c}$);
slightly red shifted; and red-asymmetric, as mentioned above. Here we
emphasized that the formula of spectral intensity Eq. (31) is obtained
by the use of the approximate formulae Eqs. (16), (21) and (24),
which are valid
only if the condition (14), e.g. $g\ll z$, $b\ll z$, is satisfied (That is,
the fractional wavelength displacement $y$ can not be indefinitely small,
$y\equiv\frac{\Delta\lambda}{\lambda_{lu}}\not\rightarrow 0$. Otherwise
$J^{c}_{y}\propto y^{-1}\rightarrow\infty $ and $k_{lu}\propto y^{-2}
\rightarrow\infty $, which are obviously unacceptable). However, the derived
formula of $I^{c}_{y}$, i.e. Eq. (31) can be safely extended to $y\rightarrow
0$ without any divergence. Particularly we have $I^{c}_{y}=0$ at $y=0$.
Therefore, Eq. (31) is valid in the whole waveband of the Cerenkov line $(0,
y_{\rm lim})$.\\

Case B. The non-uniform plane-parallel slab.

~~For a non-uniform plane-parallel slab of emissive gas, Eq. (28),
and hence Eqs.
(29)--(31) are no longer valid. In this case, the solution of the equation
of radiative transfer Eq. (27) will take the original form:
$\frac {I^{c}_{\lambda}}{n^{2}_{\lambda}}=\int^{\tau_{\lambda}(L)}_{0}
e^{-\tau_{\lambda}}\left ( \frac {J^{c}_{\lambda}}{n^{2}_{\lambda}k_{\lambda}}
\right ) d\tau_{\lambda}$. But we have pointed out that in the effecting
Cerenkov emission range, the
actual index $n_{\lambda}\simeq 1$. Therefore we have:
\begin{equation}
I^{c}_{\lambda}=\int^{\tau_{\lambda}(L)}_{0}e^{-\tau_{\lambda}}\left (
\frac {J^{c}_{\lambda}}{k_{\lambda}}\right ) d\tau_{\lambda}
\end{equation}

In the non-uniform case, the factor $(J^{c}_{\lambda}/k_{\lambda})$ can not
be taken outside the integral, because it is dependent on the position $s$,
$J^{c}_{\lambda}\propto N_{l}=N_{l}(s)$ and $k_{\lambda}\simeq k_{bf}\propto
N_{H^{0}}=N_{H^{0}}(s)$, and so is the ratio $(J^{c}_{\lambda}/k_{\lambda})$.
Therefore Eq. (34) can not be simplified to Eq. (28) (but
neutral hydrogen is an exception, for $H^{0}$, $J^{c}_{\lambda}\propto
N_{H^{0}_{l}}(s)$ and $k_{\lambda}\simeq k_{bf}\propto N_{H^{0}}(s)$, the
ratio $J^{c}_{\lambda}/k_{\lambda}$ is $s$-independent, therefore Eq.
(28)--(31) are still valid for hydrogen despite of the drastic nonuniform
variation of $N_{H^{0}}(s)$ near the front $H^{+}/H^{0}$). Eq. (34) can be
re-written as:
\begin{equation}
I^{c}_{\lambda}(L)=\int^{L}_{0}J^{c}_{\lambda}(s)e^{-\int^{s}_{0}k_{\lambda}
 ( s^{\prime} ) ds^{\prime}}ds
\end{equation}
If the effective interval where $J^{c}_{\lambda}(s)\not=0$ is $(S_{1}, S_{2})$
rather than $(0, L)$, then Eq. (35) can be rewritten as:
\begin{equation}
I^{c}_{\lambda}=\int^{S_{2}}_{S_{1}}J^{c}_{\lambda}(s)e^{-\int^{s}_{0}
k_{\lambda} ( s^{\prime} ) ds^{\prime}}ds
\end{equation}

Eq. (36) gives the emergent spectral intensity of the nonuniform plan-parallel
slab. The integral can be evaluated numerically if $J^{c}
_{\lambda}(s)$ and $k_{\lambda}(s)$ are given at each point $s$
along the ray.
However, in some cases, a semi-quantitative estimation of $I^{c}_{\lambda}$
is good enough. We suggest a simple expression with a similar form of Eq.
(30) for the uniform case, to replace Eq. (36):
\begin{equation}
I^{c}_{\lambda}=\frac {\overline{J^{c}_{\lambda}}}{\overline{k_{\lambda}}}=\frac
{\int^{S_{2}}_{S_{1}}J^{c}_{\lambda}(s)ds}{\int^{S_{2}}_{S_{1}}k_{\lambda}
(s)ds}
\end{equation}
where, the integral region $(S_{1}, S_{2})$ expresses the range where
the atom (or ion) species concerned exists, hence $J^{c}_{\lambda}(s)\not=0$,
and $k_{\lambda}(s)\not=0$.

The special importance of the non-uniform plane-parallel slab is in that,
according to the conventional photoionization model of BLR of AGNs,
most low-ionized ions particularly the hydrogenic ions, e.g.
$Fe^{+}, Mg^{+},\cdots$, exist
in a layer near to the ionization front of $H^{+}/H^{0}$, and have
a non-uniform distribution, $N_{Fe^{+}}=N_{Fe^{+}}(s)$,
$N_{Mg^{+}}=N_{Mg^{+}}(s),\cdots$.
For example, it is known that the ionization thresholds of iron are
$Fe^{0}/Fe^{+}/Fe^{++}=7.9/16.18/30.65 {\rm eV}$, and the position of
the front $H^{+}/H^{0}$ is at $\sim 13.6 {\rm eV}$. So we infer that
the low-ionized ions $Fe^{+}$ exist on both sides of the front
$H^{+}/H^{0}$, where the density
of neutral hydrogen $N_{H^{0}}$ varies drastically with $s$, $N_{H^{0}}=
N_{H^{0}}(s)$, the same as the density of $Fe^{+}, N_{Fe^{+}}=N_{Fe^{+}}(s)$.
Fig. 4 shows the schematic distribution of $Fe^{+}, H^{0}, H^{+}$, in the
neighbourhood of the front of $H^{+}/H^{0}\sim 13.6 {\rm eV}$.


\subsubsection{Extended formulae of intensity $I^{c}_{y}$ for complex
atoms(ions) gas}

~~For simplicity, we only give the formulae for the uniform and optically
thick plane-parallel emission slab. The discussion is parallel to Sect. 2.4.1
for the hydrogenic atoms. Thus Eq. (31) is still valid for the heavy elements,
only with a replacement of $N_{l}\rightarrow NS_{l}$ and $N_{u}\rightarrow
NS_{u}$(Eq.($5^{\prime}$) in the expressions of $C_{0}, C_{1}, C_{2}$
in Eqs. (32),
we then get the emergent intensity $I$ and the line profile for the
Cerenkov line of the complex atoms(ions). Furthermore, in the
optical waveband, the dominant absorber of photoionization
is still hydrogen due to the great abundance of $H$, thus $k_{bf}$ is still
given by the last equation in Eqs. (32).

However, if the Cerenkov line is located in the X-ray waveband, e.g. the
Cerenkov Fe $K_{\alpha}$ line $\sim 6.4 {\rm KeV}$, we must use
the parameters
$C_{0}, C_{1}, C_{2}$ and $k_{bf}$ given by Eq. ($20^{\prime}$),
($21^{\prime}$), $(24^{\prime\prime}$) and ($25^{\prime}$), ($25^{\prime
\prime}$) respectively. For convenience, we relist all of these parameters
in Eq. ($32^{\prime}$) below. Therefore, in Eq. (31),
\begin{displaymath}
y=\frac {\epsilon_{lu}-\epsilon}{\epsilon_{lu}}
\end{displaymath}
\begin{displaymath}
C_{0}=1.05\times 10^{-76}\epsilon^{-4}_{lu}A_{ul}g_{u}N\left (
\frac {S_{l}}{g_{l}}-\frac {S_{u}}{g_{u}}\right )
\end{displaymath}
\begin{displaymath}
C_{1}=5.77\times 10^{-53}\epsilon^{-2}_{lu}A_{ul}g_{u}N\left (
\frac {S_{l}}{g_{l}}-\frac {S_{u}}{g_{u}}\right )
\end{displaymath}
\begin{displaymath}
C_{2}=1.75\times 10^{-87}\epsilon^{-4}_{lu}A_{ul}\Gamma_{lu}g_{u}N\left (
\frac {S_{l}}{g_{l}}-\frac {S_{u}}{g_{u}}\right )~~~~~~~~~~(32^{\prime})
\end{displaymath}
\begin{displaymath}
k_{bf}=8.4\times 10^{-46}\epsilon^{-3}_{lu}N(Fe)S_{2}~~~~~~~ ({\rm for}
~~~Fe~~ K_{\alpha}-{\rm line})
\end{displaymath}
and $\epsilon_{lu}$ is in unit {\rm ergs} in CGSE system
($ 1 {\rm KeV}=1.602\times
10^{-9} {\rm erg}$).

Using Eq. (31), we calculated the Cerenkov $K_{\alpha}$ line
$\sim 6.4 {\rm KeV}$ of the iron ion $Fe^{+19}$, as shown in Fig. 5. The calculation
parameters are chosen as
$\gamma_{c}=1.0\times 10^{4}$, $N=2.5\times 10^{19}{\rm cm^{-3}}$(Fig.5 (a))
and $\gamma_{c}=1.0\times 10^{4}$, $N=2.5\times 10^{17}{\rm cm^{-3}}$
(Fig.5 (b)) respectively.


\subsection{The Cerenkov line redshift $y^{c}_{t}$ (or $\Delta Z^{c}$)}

\subsubsection{Formulae of $y^{c}_{t}$ (or $\Delta Z^{c}$) for hydrogen or
hydrogen-like gas}

~~Using Eq. (31), the small ``Cerenkov line redshift'' $y^{c}_{t}$ can be
obtained from the equation $\frac {dI^{c}_{y}}{dy}=0$. Thus we obtained:
\begin{displaymath}
\Delta Z^{c}\equiv y^{c}_{t}=\frac {\Delta\lambda^{c}_{t}}{\lambda_{lu}}
=\frac {1}{y^{-1}_{\rm lim}+\sqrt{y^{-2}_{\rm lim}+\frac{k_{bf}}{C_{2}}}}
\end{displaymath}
\begin{equation}
~~~=\frac {1}{C^{-1}_{0}\gamma^{-2}_{c}+\sqrt{C^{-2}_{0}\gamma^{-4}_{c}
+\frac{k_{bf}}{C_{2}}}}
\end{equation}

If the gas is very dense (e.g. $N_{H^{0}}\simeq 10^{16} {\rm cm^{-3}}$),
we have
a line width $y_{\rm lim}\gg\sqrt{C_{2}/k_{bf}}$ ($y_{\rm lim}\equiv \frac {\Delta
\lambda^{c}_{\rm lim}}{\lambda_{lu}}\propto C_{0}\propto N$, but $C_{2}/k_{bf}$
is density-independent, see Eqs. (20), (32). Therefore from Eqs. (32) and
(38) we get a simplified redshift formula:
\begin{displaymath}
\Delta Z^{c}\equiv y^{c}_{t}\simeq \sqrt{\frac {C_{2}}{k_{bf}}}
\end{displaymath}
\begin{equation}
~~~=1.04\times
10^{-11}\sqrt{\lambda_{lu}A_{ul}\Gamma_{lu}\frac {g_{u}}{g_{l}}R_{l}
R^{-1}_{p}p^{5}\xi}
\end{equation}
where $\xi\equiv N/N_{H^{0}}$ represents the ``abundance'' of the concerned
hydrogenic atom (ions) species. For the gas of neutral hydrogen $\xi=1$.
It is easy to show that the wavelength displacement $y^{c}_{t}$ in Eq. (39)
is just at the wavelength position where $k_{lu}=k_{bf}$. For the wavelength
region $y\leq y^{c}_{t}$ which is close to $\lambda_{lu}$,
we have $k_{lu}\gg k_{bf}$ which means that the Cerenkov line redshift is
just caused by the great line absorption $k_{lu}$ at vicinity of $\lambda_{lu}$.
From Eq. (39) we see, the redshift for the dense gas is $N$-independent,
$\Delta Z^{c}$ is only dependent on the atomic parameters ($\lambda_{lu},
A_{ul}, \Gamma_{lu},\cdots$) and the temperature $T$ of the gas, and the
``abundance'' $\xi\equiv N/N_{H^{0}}$.

Another limiting case is for the gas with lower density, which makes $y_{\rm lim}\ll\sqrt
{\frac{C_{2}}{k_{bf}}}$.  In this case (but still optically thick for continuum,
$k_{\lambda}L=(k_{lu}+k_{bf})L
\gg 1)$), Eq. (38) is simplified as:
\begin{equation}
\Delta Z^{c}\equiv y^{c}_{t}\equiv\frac {\Delta\lambda^{c}_{t}}{\lambda_{lu}}
\simeq\frac {y_{\rm lim}}{2}
\end{equation}
where, $y_{\rm lim}$ is given by Eq. (20) ($y_{\rm lim}\propto N$). The critical
point of ``higher'' and ``lower'' densities is determined by a critical
equality $y_{\rm lim}=\sqrt{\frac{C_{2}}{k_{bf}}}$, from which we get a
critical density $N^{cr.}$ of the concerned species in a gas:
\begin{displaymath}
N^{cr.}=1.54\times 10^{2}\left (\frac{g_{u}}{g_{l}}\right )^{-1/2}
\end{displaymath}
\begin{equation}
~~~\times\gamma_{c}^{-2}\lambda^{-7/2}_{lu}A^{-1/2}_{ul}\Gamma^{1/2}_{lu}p^{5/2}
(R_{l}R_{p})^{-1/2}\xi^{1/2}
\end{equation}

Therefore, Eq. (39) and Eq. (40) are valid for $N\gg N^{cr.}$ and $N\ll
N^{cr.}$ respectively. Fig. 3(c) shows the profile $I^{c}_{\lambda}\sim
\Delta\lambda$ and the redshift $\Delta\lambda^{c}$ of the Cerenkov line for
the case $N\gg N^{cr.}$. The red-asymmetry of the profile is remarkable.
But for the cases $N\simeq N^{cr.}$ and $N\ll N^{cr.}$ (see Fig. 5 (b)), the
profile becomes more symmetric. And the Cerenkov line redshift
becomes very small.

We emphasize again, both Eq. (39) for $N\gg N^{cr.}$ and Eq. (40) for
$N\ll N^{cr.}$ are derived from Eq. (31). Both Eq. (39) and Eq. (40) are
valid only for the optically thick case, $\tau_{\lambda}=(k_{lu}+k_{bf})L\gg 1$.

\subsubsection{Extended formulae of $y^{c}_{t}$ (or $\Delta Z^{c}$) for
complex atoms (ions) gas}

~~The formula of the Cerenkov line redshift, Eq. (38) is also valid for a gas
composed of complex atoms (ions). The related parameters in (38), i.e.
$C_{0}, C_{1}, C_{2}$ and $k_{bf}$ are still given by Eq. (32) only
with replacement
$\left (\frac {N_{l}}{g_{l}}-\frac {N_{u}}{g_{u}}\right )\Longrightarrow
N\left (\frac {S_{l}}{g_{l}}-\frac {S_{u}}{g_{u}}\right )$, or equivalently,
$N_{l}\rightarrow NS_{l}$, $N_{u}\rightarrow NS_{u}
$(see Eq.($5^{\prime\prime}$),
with unchanged $k_{bf}$ because the dominant photoionization absorber in the
optical wavelength is still hydrogen.


In parallel, in the limiting cases,
$y_{\rm lim}\gg\sqrt{\frac{C_{2}}{k_{bf}}}$ and
$y_{\rm lim}\ll\sqrt{\frac{C_{2}}{k_{bf}}}$,
or equivalently, $N\gg N^{cr.}$ and
$N\ll N^{cr.}$, the simplified redshift formulae are still given by Eq. (39)
and (40) respectively.

If the Cerenkov line of complex atoms or ions is in the X-ray band, it is
better to change the wavelength in (32) into the photon energy,
$\lambda_{lu}\rightarrow\epsilon_{lu}$, $\lambda_{lu}=\frac {hc}{\epsilon_{lu}}$,
i.e. the parameters $C_{0}, C_{2}$ and $k_{bf}$ in (38) have to
be expressed by Eq. ($32^{\prime}$).
given in Sect. 2.4.2, where $\epsilon_{lu}\equiv\epsilon_{u}-\epsilon_{l}
=h\nu_{lu}$ is in unit {\rm ergs} in the CGSE system, rather than
{\rm KeV} (1{\rm KeV}=$1.602
\times 10^{-9} {\rm ergs}$).

\subsection{The emergent intensity $I^{c}$ of the Cerenkov line}

\subsubsection{Formulae of $I^{c}$ for hydrogen of hydrogen-like gas}

~~Integrating Eq. (31), we obtained the total intensity $I^{c}$ of the Cerenkov
line, for the optically thich uniform plane-parallel layer $I^{c}=\int_{0}^{y_{\rm lim}}I^{c}_{y}dy$
Thus
\begin{equation}
I^{c}=Y\left [ \ln {(1+X^{2})}-2\left ( 1-\frac {\arctan{X}}{X}
\right )\right ]
\end{equation}
where, $Y\equiv\frac{N_{e}}{2}\frac {C_{1}}{k_{bf}}$, $N_{e}$ is the density
of relativistic electrons, $X\equiv\sqrt{\frac {k_{bf}}{C_{2}}}y_{\rm lim}$,
$C_{0}$, $C_{1}$, $C_{2}$ and $k_{bf}$ are given by Eq. (32), $y_{\rm lim}$ is
given by Eq. (20).

Eq. (42) is in principle valid for a uniform plane-parallel slab
of emissive gas. But as an exception, it can be used to
calculate the line intensity of the neutral hydrogen without any problem.
Although the density of $H^{0}$ atoms, $N_{H^{0}}$ varies with depth $s$
drastically near the front $H^{+}/H^{0}$, $N_{H^{0}}=N_{H^{0}}(s)$, we
point out that the ratio $(J^{c}_{\lambda}/k_{\lambda})_{H^{0}}$ is constant,
despite the nonuniform ionization structure in the neighbourhood of the
front $H^{+}/H^{0}$. Therefore $(J^{c}_{\lambda}/k_{\lambda})_{H^{0}}$ can
be taken out of the integral $I^{c}_{\lambda}=\int_{0}^{\tau_{\lambda}(u)}
e^{-\tau_{\lambda}}\left (\frac{J^{c}_{\lambda}}{k_{\lambda}}\right )d\tau
_{\lambda}\simeq \frac{J^{c}_{\lambda}}{k_{\lambda}}$, which ensures the
validity of Eq. (42).

For the non-uniform case, the low-ionization line, such as $Mg^{+}, Fe^{+}$,
$O^{+},\cdots$, the emergent intensity of Cerenkov line $I^{c}$, is also
obtained from  $I^{c}=\int_{0}^{y_{\rm lim}}I^{c}_{y}dy$, and
$I^{c}_{y}$ is given by Eq. (35). The calculation is somewhat complicate.
However, in the semi-quantitative estimation of $I^{c}$, we can still invoke
Eq. (42) only regarding the factors $R_{l}\equiv N_{l}/N$,
$R_{p}\equiv N_{H^{0}_{p}}/N_{H^{0}}$, particularly, $\xi\equiv N/N_{H^{0}}$
etc. as the average value $\overline{R_{l}}, \overline{R_{p}}, \overline{\xi},
 \cdots $,
in the interval $(S_{1}, S_{2})$, where the concerned atoms or ions exist.

\subsubsection{Extended formulae of $I^{c}$ for complex atoms (ions ) gas}

~~The emergent total intensity of Cerenkov line of the complex atoms (ions) in
optically thick case is also given by Eq. (42), where $C_{0}, C_{1}, C_{2}$
and $k_{bf}$ are given by Eq. (32) replacing
$N_{l}$ by $NS_{l}$ and $N_{u}$ by $NS_{u}$. Note that,
when the Cerenkov line is in the X-ray band, Eq. ($32^{\prime}$) has to be
used to replace Eq. (32).
\section{Conclusion and Discussion}
~~The Cerenkov line-like emission of the relativistic electrons, passing
through an optically thick dense gas, which we suggested early in 1980, has
been verified by elegant laboratory experiments (Xu et al. 1981, 1988, 1989). In
 this paper, we give a detailed and clearer physical discussion and
emphasize the potential importance of this new mechanism for  high-energy
astrophysics, and give the extended and improved formula system describing
the emergent intensity, the line profile, the line width and the Cerenkov
redshift of the Cerenkov line, among which the extension of formulae to the
multi-electron complex atoms(ions) has special significance for the study
of the broad lines of heavy elements in AGNs, particularly for lines in the
X-ray band.

A possible application of the new line emission mechanism is in the
exploration of the origin of the broad $\sim 6.4 {\rm KeV}$ $K_{\alpha}$
line of the low-ionized iron ions of Seyfert1 galaxies. Now the disk-line
models, in which the $\sim 6.4 {\rm KeV}$ $K_{\alpha}$ line is regarded as
one of the reflection components from the disk, strongly illuminated by the
hard X-ray continuum, are widely accepted to explain the origin and
characters of the broad $K_{\alpha}$-line with asymmetric profile. It is
believed that the ``Compton reflection and iron fluorescence
features'', provide a powerful probe of the accretion flow and the strong
gravitational field (see e.g. Georgy \& Fabian 1991; Reynolds, Fabian \&
Inoue 1995). However, people now question the relativistically smeared
disk-line interpretation. According to the disk-line model, both the
line profile and the position of the peak are  dependent on the inclination
angle $\theta$ of the disk, and therefore are different from one sample to anoth
er.
But the observations show similar profiles of the $K_{\alpha}$ line for all
Seyfert1 galaxies, even for Seyfert2 galaxies, with nearly unchanged peak at
$\sim 6.4 {\rm KeV}$(Nandra, 1997,a,b), which implies that the $K_{\alpha}$
line might not be from the inner disk, as thought before. It seems more
reasonable to replace the ``old disk'' by  ``cold cloudlets'' and/or 
``cold filaments'' around the central massive black hole. Furthermore, the line
emission mechanism might not, or rather, not only be 
photoionization-fluorescence. The photoionization-fluorescence
model predicts positive correlations of both the light
curves and the fluxes between the $K_{\alpha}$ line and the X-ray continuum.
But the observations do not confirm this (e.g. Lee et al.  1999). Besides, the p
rediction of a marked absorption
dip at edge $>7{\rm KeV}$ which always accompanies the fluorescent
$\sim 6.4 {\rm KeV}$ $K_{\alpha}$ line is also not confirmed by the
observations(Young et al. 1998).

We have shown that  fluorescence is not a  `unique'
mechanism of the line-emission of the low-ionization iron in the X-ray band.
Another line mechanism which can produce the $\sim 6.4 {\rm KeV}$
$K_{\alpha}$ line is the Cerenkov line-like emission, as described in this
paper. For a very dense gas, optically thick for the continuum, the Cerenkov
line becomes the unique emission line which can escape from the surface of
the cloud of dense gas. The Cerenkov line will be strong to match the
observation when the density of relativistic electrons in BLR is high enough.
Therefore this kind of line emission might be a new possible
mechanism to attack the
$6.4{\rm KeV}$ $K_{\alpha}$ problem of AGNs. We expect that some puzzles of
$K_{\alpha}$ line of iron could be resolved in  this way(e.g. the
observed strange correlations of the light curves and the fluxes between the
$K_{\alpha}$ line and X-ray continuum radiation), even though there remain a
lot of problems to be solved.

\acknowledgements {This research is supported by the ``National Foundation
of Nature Science'' and ``National Pandeng Plan of China''. }

{}

\newpage
\begin{figure}
\centerline{\psfig{figure=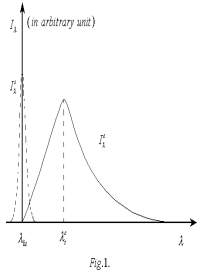,width=4.0cm,height=4.0cm}}
\caption{Comparison of the schematic profiles of the Cerenkov line $I^{c}_{\lambda}\sim\lambda$
and the normal line by spontaneous transition process $I^{s}_{\lambda}\sim\lambda$.
The former has been red-shifted and has asymmetry and large linewidth.}
\end{figure}

\begin{figure}
\centerline{\psfig{figure=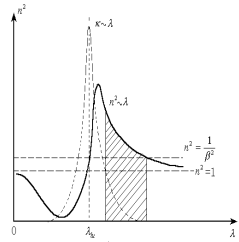,width=4.0cm,height=4.0cm}}
\caption{Relation between refractive index $n$ and wavelength $\lambda$,
 and relation between extinction coefficient and wavelength. The Cerenkov radiation survives
in the shaded narrow region.}
\end{figure}

\begin{figure}
\centerline{\psfig{figure=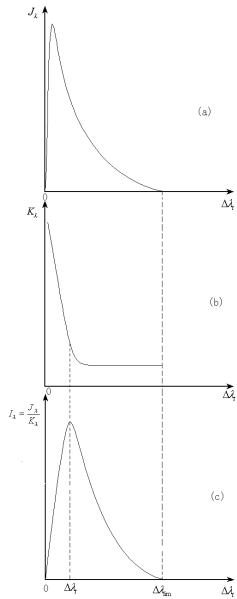,width=4.0cm,height=8.0cm}}
\caption{The Cerenkov emissivity $J_{\lambda}$ (a), absorption coefficient
  $k _{\lambda}$ (b), and the emergent intensity $I_{\lambda}$ of the Cerenkov line
  (optically thick case) (c). From (c) we see, the Cerenkov linewidth is large. The line profile is      asymmetric with a small redshift.}
\end{figure}

\begin{figure}
\centerline{\psfig{figure=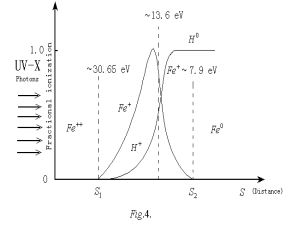,width=4.0cm,height=4.0cm}}
\caption{The schematic distributions of $Fe^{+}~and ~H^{0}$ near the $H^{+}/H^{0}$ ionization front,  i.e. the ionization structure of the surface layer of the cloud in BLR.}
\end{figure}

\begin{figure}
\centerline{\psfig{figure=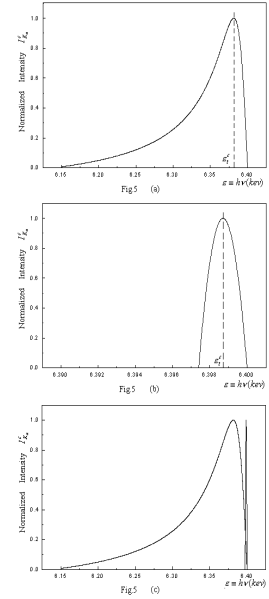,width=4.0cm,height=8.0cm}}
\caption{The emergent  intensity $I^{c}_{\lambda}$ of $Fe^{+19}$ ions.
(a) shows the line profile of
$Fe^{+19}$ in gas of higher density $\gamma_{c}=1.0\times 10^{4}$,
$N=2.5\times 10^{19}$; (b) shows
a symmetric profile in gas of lower density $\gamma_{c}=1.0\times 10^{4}$,
$N=2.5\times 10^{17}$;
(c) combines the line-profiles of (a) and (b) for comparason.}
\end{figure}

\end{document}